# Differences in cell death and division rules can alter tissue rigidity and fluidization

Gudur Ashrith Reddy [1,2] and Parag Katira* [1,3]

[1] Mechanical Engineering Department, San Diego State University, San Diego, CA, USA
[2] Department of Bioengineering, University of California – San Diego, San Diego, CA, USA
[3] Computational Science Research Center, San Diego State University, San Diego, CA, USA
* Corresponding Author, email address: pkatira@sdsu.edu

## Abstract

Tissue mechanical properties such as rigidity and fluidity, and changes in these properties driven by jamming-unjamming transitions (UJT), have come under recent highlight as mechanical markers of health and disease in various biological processes including cancer. However, most analysis of these mechanical properties and UJT have sidestepped the effect of cellular death and division in these systems. Cellular apoptosis (programmed cell death) and mitosis (cell division) can drive significant changes in tissue properties. The balance between the two is crucial in maintaining tissue function, and an imbalance between the two is seen in situations such as cancer progression, wound healing and necrosis. In this work we investigate the impact of cell death and division on tissue mechanical properties, by incorporating specific mechanosensitive triggers of cell death and division based on the size and geometry of the cell within *in silico* models of tissue dynamics. Specifically, we look at cell migration, tissue response to external stress, tissue extrusion propensity and self-organization of different cell types within the tissue, as a function of cell death and division and the rules that trigger these events. We find that not only do cell death and division events significantly alter tissue mechanics when compared to systems without these events, but that the choice of triggers driving these cell death and division events also alter the predicted tissue mechanics and overall system behavior.

## Introduction

In dense tissue constructs, with cell packing fractions close to 1, the ability of cells to migrate and exchange neighbors dictates whether the tissue is classified as "Jammed" or "Unjammed"[1–3]. Systems with low cell mobility and none to a few neighbor exchanges in timescales spanning hours, are classified as jammed. On the other hand, increased cell mobility and neighbor exchanges within this time scale are characteristics of an unjammed tissue[4]. Intermediate scenarios where certain cells rapidly exchange neighbors, or where a small region of the tissue shows localized jamming and unjamming are also observed[5]. The state of cells being jammed or unjammed also relates to the bulk tissue being mechanically classified as being rigid or fluid, where jammed tissues are classified as rigid while unjammed tissues are classified as fluid[6]. However, at this point, the definitions can get a bit blurry. Tissue rigidity and fluidity might refer to the ability of the bulk tissue to deform and rearrange under external stresses and dissipate these stresses across the structure. While jamming and unjamming of individual cells will contribute to these bulk level outcomes, migration of cellular collectives within the tissue, with few individual neighbor exchanges, may also drive tissue reorganization and fluidization[7]. Ultimately, mechanical characterization of tissues as being rigid or fluid, and jammed or unjammed, can help describe and understand different mechanical states of these tissue systems in a variety of biological processes such as embryogenesis, tissue regeneration, wound healing, and diseases such as cancer.

The above described mechanical properties of a dense tissue are usually known to be dictated by three individual cell level properties – 1) the migration ability of individual cells[8], governed by intracellular force generation and transmission of these forces from the cell to the surrounding tissue, 2) the persistence of cell migration[9], driven by the fluctuations in the directions of aforementioned forces, and 3) a target cell shape index[7,10], dictated by the mechanical stiffness of the cell cytoskeletal cortex and the adhesion between neighboring cells. These factors and their effect on tissue mechanical states have been extensively studied and described in literature[1,11–13]. However, there are additional factors such as inter-cellular heterogeneity within the tissue[5,14], formation of supra-cellular structures spanning multiple cells[15,16], mechanical properties of the cell nucleus[17], and extensile vs. contractile nature of cellular collectives[18] that have also been shown to influence the mechanical state of these tissues. One specific factor, cellular death and division within the tissues, while implicated in broader tissue level dynamics, has been largely left out of consideration when discussing tissue mechanics such as tissue rigidity, fluidity, jamming and unjamming. A potential reason behind this might be the intuitive assumption that cell death and division will create large enough changes in the local cell density that can instantaneously drive tissue fluidization and unjamming. However, we hypothesize that not only the inclusion of cell death and division, but also the choice of triggers that cause cellular death and division can have a significant impact on the mechanical states of tissues. Cell death and division processes are highly mechanosensitive in nature and strongly depend on a number of key cellular parameters such as cell size[19,20], local cell density[21], cell shape[22,23], interface area between cells[24], intra-cellular tension[25], and paracrine biochemical signaling[26]. Many of these cellular parameters such as size, shape, interface area, and intra-cellular tension are related to the three individual cell level properties – migration speed, persistence and shape index, known to drive jamming-unjamming and rigidity-fluidity transitions in tissues. Thus, not considering cellular death and division triggered via mechanosensitive pathways while describing tissue mechanics during biological properties such as embryogenesis, wound healing and cancer progression limits our understanding of these processes.

Self-propelled Voronoi (SPV) models are a commonly used computational tool to study how cell-level interactions influence tissue mechanical properties and outcomes[27]. We use this modeling framework to examine how cell death and division influences tissue mechanical properties. Our focus is specifically on how the choice of death and division triggers – either random, based on cell size, based on cell shape, or based on cell interface length with other cells (which combines both cell size and shape), influences tissue mechanics. Each of these triggers has a specific biological relevance and can be the dominant trigger for cell-death and division events in specific environments. For example, cell size is known to strongly control cell division likelihood. Also, cell size is related to local cell density, which has been shown to govern cell proliferation rates in growing tissues systems. Cell shape on the other hand can drive key biochemical signaling pathways as well as nuclear localization of key transcription factors driving cell fate changes. Cell interface length with other cells, which is quantified by the cell perimeter, considers inter-cellular signaling via cell adhesion junctions which are mechanosensitive triggers for cell fate decisions. It also takes into account cellular stretch, driven by both size and shape changes in the cell, which can alter intra-cellular tension. In this work, we show that while death and division events indeed alter tissue mechanics by driving unjamming transitions, the extent of unjamming depends on the cell death and division triggering mechanism. Additionally, we show that local unjamming may not directly relate to tissue fluidization as defined by the ability of a tissue system to dissipate external stresses. Overall, this work highlights the importance of properly considering the rules governing cellular death and division while making predictions on tissue mechanical properties and their role in governing outcomes of key biological properties.

## Methods

### Simulation Setup

We simulated the cells as Voronoi polygons[7,28] (Figure 1 A) using MATLAB. In all configurations we start with an initial number of cells equal to 900 in a square shaped region with an area of 9x10$^{-8}$ $m^2$ such that the cells have an average area of 10$^{-10}$ $m^2$. The energy of the tissue is calculated as a function of the cell cortex stiffness, inter-cellular adhesion, and the internal pressure of the cell[10,29,30]. This energy is given by[31] –

$$E_i = \frac{1}{2}k(P_i - P_0)^2 - \frac{1}{2}\sum_j \gamma_j l_i - \lambda \log \frac{A_i}{A_0} \tag{1}$$

where $k$ is the cell cortex stiffness and $P_i$ and $P_0$ are the perimeters of the cell as a polygon and of a circle with the same area as the cell. $\gamma_j$ is the lower of the two adhesion coefficients at the interface of contact between the cells and $l_i$ is the length of the interface. $\lambda$ is the internal pressure parameter and $A_i$ and $A_o$ are areas of the cell and the mean area available to each cell (total area available to all cells divided by the number of cells) respectively.

The simulation progresses through cellular rearrangements, where ~1% of the cells are randomly chosen and moved a distance which is exponentially distributed with a mean of 0.5 $\mu m$, corresponding to ~5% of the length of the cell, in a random direction. If this interim configuration has a lower energy, the configuration is always accepted and configurations which increase the energy of the tissue are accepted with a probability based on the Metropolis algorithm expressed as $e^{\frac{E-E_i}{k_B T}}$ where $E$ and $E_i$ are the current and interim energies of the tissue respectively and $k_B T$ is an effective temperature term that corresponds to internal energy the cells can generate for migration[32,33].

In all our simulations, we have an arbitrary unit of time which is the number of configuration acceptances. We convert that into units of time so we get a better sense of the variation of the properties. The average distance a cell can move in any iteration is 0.5 $\mu m$. Using an average cell speed of 20 $\mu m/h$ from previous studies[34], we approximate that one configuration acceptance corresponds to approximately 1.5 minutes.

**Initial Setup**

Prior to running our simulations with cell death and division, we must arrive at an initial tissue configuration upon which further analysis can be done. Additionally, we require some properties of the cells in the tissue such as the mean area, perimeter, and shape. These mean values - $A_m$, $P_m$ and $S_m$ describe the properties which the cells are energetically favorable to acquire in the absence of death and division and will play a significant role in the dynamics of the tissue. This is done by starting our simulation with cells randomly distributed in the available space, followed by running the Monte Carlo algorithm described above. The cells rearrange to minimize the energy in the absence of death and division. Once the energy of the tissue has stabilized, we consider it to be optimized and proceed to extract the mean values of the properties described above and the final coordinates of the centers of the cell which is used as the starting configuration for the simulations to follow.

**Death and Division**

All cells are assigned characteristic death and division probabilities based on which of the four rules is under consideration. For the random rule all the cells have equal probabilities for death or division, whereas for the area, perimeter and shape rule the probabilities are defined to be cell morphology specific - more elongated cells are assigned a higher likelihood of dividing and a lower chance of dying, while more rounded, compact cells are assigned higher likelihood of death and a low chance of division based on the equations (2-4) below –

$$p_{death,1} = p^{\frac{X_i - X_0}{X_m - X_0}} \tag{2}$$

$$p_{death,2} = \frac{1}{1 + (\frac{1}{p} - 1)^{\frac{1}{p}\frac{X_i - X_m}{X_0}}} \tag{3}$$

$$p_{division,1} = \frac{1}{1 + (\frac{1}{p} - 1)^{\frac{1}{p}\frac{-(X_i - X_m)}{X_0}}} \tag{4}$$

where $p$ is the baseline probability which dictates the likelihood that the average cell undergoes death or division. $X_i$, $X_0$ and $X_m$ are the current properties of the cell (cell area, perimeter or shape), the cell property if the cell was a circle (the rest state of a free cell suspended in solution) and the mean properties of all the cells extracted from the initial optimization where $X$ may be the area, perimeter, or shape (perimeter divided by the square root of area) of the cell. We propose two different equations for the probability of death to accommodate the differences in the initial distributions of the area, perimeter and shape of the cells.

For the random death and division rule, the probabilities are simply equal to the baseline probability $p$. For the other rules, we propose two possible $p_{death}$ equations which differ in their sensitivity to the property $X_i$ (Figure 1). Our choice on which of the two probability equations to use for each of the rules is based on the fact that the tissue must reach homeostasis. There are two considerations for this – firstly, under our division mechanism the area is perfectly conserved whereas the perimeter and shape are not. This means that the daughter cells have exactly half the area of the parent cell, but will have a perimeter and shape of greater than half of the original cell. Secondly, when we plot the distribution of the area, perimeter and shape of the cells after the initial optimization (Figure 1 C-E) we find that the perimeter and shape of the cells have more of a positive skew as compared to the area rule. They have kurtosis values of 17.46, 5.31 and 3.46 for area, perimeter and shape respectively meaning the perimeter and shape distributions have much fatter tails and a larger proportion of values greater than the median. Based on this, we use $p_{death,1}$, which is the steeper of the two equations for the perimeter and shape rules and use $p_{death,2}$ for the area rule. Additionally, there is a nuance for the shape-based probability equation which relates to cells with very high shape values. The equations currently state that as the shape of the cell increases over the mean shape, there is higher likelihood for the cell to divide. However, we notice that for large values of the cell shape ($S_i > 6$) the cell is extremely elongated or stretched out and is not likely to divide anymore. Hence, we apply the equation for $S_i \leq 6$ only. These equations are summarized in Table 1.

Table 1: Equations describing death and division probabilities at the cellular level

| Rule | $p_{death}$ | $p_{division}$ |
|---|---|---|
| Area | $\dfrac{1}{1 + (\frac{1}{p} - 1)\frac{1}{p}^{\frac{A_i - A_m}{A_0}}}$ | $\dfrac{1}{1 + (\frac{1}{p} - 1)\frac{1}{p}^{\frac{-(A_i - A_m)}{A_0}}}$ |
| Perimeter | $\min(1, p^{\frac{P_i - P_0}{P_m - P_0}})$ | $\dfrac{1}{1 + (\frac{1}{p} - 1)\frac{1}{p}^{\frac{-(P_i - P_m)}{P_0}}}$ |
| Random | $p$ | $p$ |
| Shape | $p^{\frac{S_i - S_0}{S_m - S_0}}; S_i < 6$ | $\dfrac{1}{1 + (\frac{1}{p} - 1)\frac{1}{p}^{\frac{-(S_i - S_m)}{S_0}}}; S_i < 6$ |

Once a cell is marked for death or division, we change the mechanical properties of the cell for a short duration of time to mimic the forces which dying or dividing cells apply on their adjacent cells. Dying cells are given a negative internal pressure coefficient so that it is energetically favorable for them to shrink in size before death[35], and dividing cells have a much higher stiffness and lower adhesion so that they round up before division[36,37]. Cell death is simulated by deleting the point corresponding to the center of the cell. Cell division is simulated by replacing a dividing cell by two daughter cells placed equidistantly along the longest diagonal of the mother cell. We assume that the daughter cells inherit the mechanical phenotype of the mother cells. Cell death and division events are interspersed between cell rearrangement iterations. Furthermore, we notice that the daughter cells from a newly divided cell are likely to be smaller than the average cell which would in turn increase their likelihood to die. To avoid this, we prevent the daughter cells from dying for the duration of at least two death and division cycles. This introduces dynamic features into the model as cells continuously change their geometrical properties, which influences their likelihood for division or death.

**Velocity and Displacement Measurements**

By tracking the movement of the centers of the cells over time, we can arrive at two important metrics to better understand our results – the instantaneous velocity and mean square displacement. The instantaneous velocity is obtained by calculating the mean displacement of all the cells over 100 configuration acceptances, with the cells dying and dividing during this interval excluded from the calculation. The mean square displacement (MSD) is calculated in a similar manner, and is used to quantify the diffusive propensity of the cells by fitting the obtained data with the diffusion equation below –

$$MSD = K_\alpha t^\alpha \tag{5}$$

The constants $K_\alpha$ and $\alpha$ describe the non-linearity of the relationship between MSD and time. Furthermore, we calculate these constants for different global time intervals, to compare the nature of early and late timescale events under the same duration of observation.

**Tissue Mechanical Properties**

In order to extract the mechanical properties of the tissue as a whole, we subject the tissue to constant strain rates ranging from $0.1\ nm/Iter$ to $40\ nm/Iter$ implemented as moving the left boundary wall of the tissue at the corresponding rate so as to compress the tissue. We model the tissues as viscoelastic materials[38,39]. We then extract the evolution of energy in the tissue and fit it to one of 5 viscoelastic models. These models are the Maxwell model, the Kelvin representation of the standard linear solid model and some variations of the same where we interchange the springs and dashpots and also add some contractile/extensile elements which exert a constant force. The energy expressions for these material systems are summarized in Table 2 (Derivation in Supplementary Table 3).

Table 2: Equations describing energy of tissue as function of the applied strain rate and the tissue stiffness and viscosity

| | Model Diagram | Energy Expression |
|---|---|---|
| 1 | 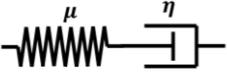 | $E = \dfrac{\mu}{2}(\dfrac{\eta\dot{\varepsilon}}{\mu}(e^{\frac{-\mu t}{\eta}}-1))^2$ |
| 2 | 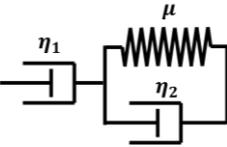 | $E = \dfrac{\mu}{2}(\dfrac{\eta_1\dot{\varepsilon}}{\mu}(1-e^{\frac{-\mu t}{\eta_1+\eta_2}}))^2$ |
| 3 | 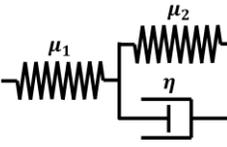 | $E = \dfrac{\mu_1}{2}(\dot{\varepsilon}t - \dfrac{\eta\mu_1\dot{\varepsilon}}{(\mu_1+\mu_2)^2}\left(e^{\frac{-(\mu_1+\mu_2)t}{\eta}}-1\right) - \dfrac{\mu_1\dot{\varepsilon}t}{\mu_1+\mu_2})^2$ $+ \dfrac{\mu_2}{2}(\dfrac{\eta\mu_1\dot{\varepsilon}}{(\mu_1+\mu_2)^2}\left(e^{\frac{-(\mu_1+\mu_2)t}{\eta}}-1\right) + \dfrac{\mu_1\dot{\varepsilon}t}{\mu_1+\mu_2})^2$ |
| 4 | 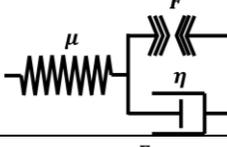 | $E = \dfrac{\mu}{2}((\dfrac{\eta\dot{\varepsilon}}{\mu}+\dfrac{F}{\mu})(e^{\frac{-\mu t}{\eta}}-1))^2$ |
| 5 | 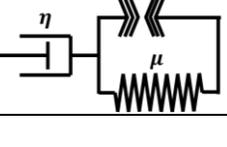 | $E = \dfrac{\mu}{2}((\dfrac{\eta\dot{\varepsilon}}{\mu}-\dfrac{F}{\mu})(1-e^{\frac{-\mu t}{\eta}}))^2$ |

Where $\mu_i$ is the stiffness of the spring, $\eta_i$ is the damping coefficient of the dashpot, $F$ is the constant force in the contractile element and $\dot{\varepsilon}$ is the constant strain rate applied to the tissue. Lastly, we compute the relaxation time using –

$$\tau_r = \dfrac{\eta}{\mu} \qquad (6)$$

**Expansion Configuration**

We would like to see how the tissue expands into an empty space, under each of the rules. To investigate this, we construct a channel with a width 1/3rd of the boundary and look at what rate the leading cell travels into the channel. We initially fill the channel with phantom cells which do not have any stiffness or adhesion but have a negative internal pressure which makes it energetically favorable for those cells to shrink in size and allow the normal cells to fill into the channel.

**De-Mixing and Mixing Between Different Cell Types**

In earlier studies we have seen that cell death and division is crucial in determining the extent to which tissues with two different types of cells segregate or de-mix. It has been shown that without death and division, the tissue fails to de-mix. Here we investigate the extent to which the tissues segregate under each of the death and division rule. We consider two populations – one with the baseline stiffness and the other with half of the stiffness, and two starting configurations – one with the cells completely mixed initially and the other where the cells are completely segregated such that the stiffer cells are on one side of the tissue and the softer cells on the other. We then run the simulation and use the segregation parameter and interface length to quantify the extent of de-mixing or interpenetration in each of the tissues.

## Results

**Initial Optimization**

After running the simulation for $10^6$ iterations, we consider the tissue to be optimized since the energy decreases by less than 0.5% over $10^3$ iterations. Figure 1(B) shows this energy minimization. The mean area, perimeter and shape values are 1.057 x $10^{-10}$ m$^2$, 4.411 x $10^{-5}$ m and 4.297 respectively. This optimization ensures that the immediate succeeding changes to the initial configuration are driven majorly by death and division since the energy from rearrangements has already been minimized.

**Homeostasis**

The primary requirement for the tissues under each of the death and division rules is that they reach homeostasis over time. This means that we must ensure the population does not grow indefinitely or decrease to zero. This is necessary to check the validity of the death and division probability equations which we have arrived at empirically. Figure 2(A) shows us that under each of the rules, the population does not go outside ~10% of the initial number of cells over the course of 5 x $10^5$ iterations (> 1,000 hours), which satisfies the homeostasis condition.

Although all the tissues reach homeostasis, we see that the variability of the number of cells at short intervals of time is drastically different. We hypothesize that this is due to difference in turnover rates of the tissues. Looking at Figure 2(B), we see that the perimeter rule has the highest turnover rate, followed by area, random and lastly the shape rule. The death and division equations we propose dictate that more fat tailed distributions of the cell properties will have a higher turnover rate because a higher proportion of cells will have high death and division probabilities. This is confirmed from Figure 1(C) and by calculating the excess kurtosis, which is 0.89, 0.35 and 0.98 for the area, perimeter and shape rule respectively. Since a lower excess kurtosis corresponds to a distribution with fatter tails, the perimeter rule generates the tissue with the highest inter-cellular variability in probabilities. This could have important implications for the tissues when we look at the mechanical properties and mobility of the cell.

Figure 2(D) shows that the perimeter and area rule have the fastest cells, followed by the random and shape rule with the cells in the tissue without death and division the slowest. However, Figure 2(C) shows that the random rule has cells with the highest diffusivity. This means that although the random-based cells are slower, they have a higher persistence length compared to the other rules. The cells undergo sub-diffusion under all the rules since $\alpha < 1$ as seen in Figure 2(E). The stability of $K_\alpha$ and $\alpha$ over different global time intervals (Figure 2(E-F)) is additional evidence of homeostasis in these tissues.

**Response to Mechanical Compression**

Mechanical properties of a tissue constitute its stiffness and viscosity. Traditional methods to measure the mechanical properties of a tissue include mechanical testing[40], imaging techniques such as magnetic resonance elastography[41] and ultrasonography[42], and probing techniques such as atomic force microscopy[43]. Here, we use a variation of uniaxial compression tests where we subject the tissue to a constant strain rate as opposed to a constant stress.

We fit the energy profiles of the tissue samples to the restorative energy equations (Table 1) using the trust region reflective algorithm. We use the adjusted $R^2$ and the bounds of the coefficients to judge the validity of the fit. Models which have a high $R^2$ and coefficients with bounds the same order of magnitude of the coefficient were accepted. From Supplementary Table 1-2 we see that most of the models have a valid $R^2$ but with very large confidence intervals for the coefficients. This indicates that the 3-element viscoelastic models (Models 2, 4 and 5) overfit the data, and a 2-element model is sufficient. Of the models which meet the criteria, we see that the Maxwell model is the only model which fits data across all the rules. Interestingly, in some cases the 3-element model - Model 3 fits the data very well with reasonable bounds for the parameters. However, upon closer inspection we see that this is because Model 3 collapses from a 3-element model to the 2-element Maxwell model (Model 1) with $\mu_2 \simeq 0$ and $\mu_1 = \mu$ which further justifies our model choice.

Qualitatively, we see that the energy for the area and perimeter rule saturate and for the random and shape rule continually increase (Supplementary Figure 1). The stiffness, viscosity and relaxation time of the tissues are summarized in Table 3.

Using a characteristic length of 10 μm (length of cell), we convert the obtained values of $\mu$, $\eta$ and $\tau_r$ into standard units and get that they lie in the order $10^1 - 10^4 \, Pa$, $10^5 - 10^7 \, Pa.min$ and $10^1 - 10^2 \, hours$ respectively. We see that these parameters are of the same order of magnitude as seen in previous animal studies for each of the stiffness[44–46], viscosity[47] and relaxation time[46,48].

Table 3: Mechanical properties of the tissues under each of the death and division rules for $\dot{\varepsilon} = 2 \, nm/Iter$

| | | Area | Perimeter | Random | Shape | Without Death and Division |
|---|---|---|---|---|---|---|
| Baseline Stiffness (200 Pa) | $\mu \, (\frac{pN}{nm})$ | 13.5 ± 4.0 | 264.8 ± 53.4 | 1.1 ± 0.3 | 1.5 ± 0.0 | 1 ± 0.0 |
| | $\eta \, (\frac{\mu N.Iter}{nm})$ | 14.4 ± 2.0 | 79.2 ± 8.7 | 21.6 ± 2.1 | 30.7 ± 1.2 | 18.0 ± 0.2 |
| | $\tau_r \, (Iter)$ | 1130 ± 213 | 306 ± 32 | 19798 ± 3977 | 20188 ± 1139 | 17705 ± 188 |
| Half Stiffness (100 Pa) | $\mu \, (\frac{pN}{nm})$ | 9.6 ± 2.9 | 214.9 ± 32.2 | 1.0 ± 0.0 | 1.3 ± 0.2 | 0.9 ± 0.0 |
| | $\eta \, (\frac{\mu N.Iter}{nm})$ | 11.9 ± 1.4 | 73.4 ± 7.5 | 19.7 ± 1.0 | 33.3 ± 2.5 | 18.2 ± 0.3 |
| | $\tau_r \, (Iter)$ | 1310 ± 214 | 344 ± 17 | 19091 ± 1194 | 25521 ± 4893 | 19483 ± 510 |

All the rules generate a higher stiffness and viscosity than the tissue with no death or division[10,49,50], with the random rule only marginally higher. The relaxation time $\tau_r$ is a measure of the extent of fluidity in a material. We see that the cell size-dependent rules, perimeter and area produce much more fluid tissues than the other rules (Figure 3(E)). As expected, all the tissues demonstrate an increase in stiffness and viscosity with increasing strain rate (Figure 3(C-D)), but at very different rates. At high values of strain rate, all tissues begin demonstrating fluid-like characteristics because the external stimulus is sufficient to cause neighbor exchanges between the cells[7]. Increased cell density (Figure 3(B)) is a hallmark of tissue jamming[51], which is demonstrated in the random, shape and no death and division rules. Also, these more jammed tissues show a transition from more solid-like behavior to fluid-like response[52,53] as seen by the abrupt change in the slope of the relaxation time curve around $\dot{\varepsilon} = 1 \, nm/Iter$ (Figure 3(E)), whereas the perimeter and area rule show a more gradual change. Lastly, the properties of the tissues do not change much when we use cells with a lower stiffness, which indicates that the nature of death and division is much more important than the individual cell properties in determining tissue mechanics.

**Expansion Propensity**

From Figure 4(B) we see that some rules cause the cells to be much more invasive than others. Early in the simulation, all the rules demonstrate similar invasiveness, but at larger times the area and perimeter rules have their leading cell fluctuating near the boundary whereas the random and shape rules[54] cause the leading cell to continuously travel into the channel.

This is surprising because we would expect the more fluid tissues of area and perimeter to have a higher invasiveness. We postulate that this is because of the size dependence on death and division probabilities of the area and perimeter rules, and the lack of adhesion with the phantom cells driving changes in the energy calculation (Equation 1) that make interfaces between actual and phantom interfaces less favorable.

Table 4: Distance travelled by leading cell into the channel

|  | Area | Perimeter | Random | Shape | Without Death and Division |
|---|---|---|---|---|---|
| Distance Travelled by Leading Cell (μm) | 23.9 ± 1.3 | 20.9 ± 2.5 | 150.2 ± 55.0 | 69.6 ± 4.5 | 15.6 ± 1.5 |
| Area Occupied by Invading Cells (μm²) | 412 ± 194 | 292 ± 4 | 5265 ± 2666 | 3470 ± 111 | 210 ± 82 |

**Mixing and De-mixing of Cells with Different Stiffnesses**

In line with previous studies, we see that death and division, irrespective of the rule, drives the separation of stiff and soft cells into distinct phases[12]. We quantify the segregation by calculating the segregation parameter, which is defined as the total number of unlike neighbors divided by the total number of neighbors. A higher segregation parameter corresponds to a more mixed system. We were able to reproduce past results (Supplementary Figure 2) where we see that a lack of death and division does not cause segregation in the tissue. However, we see that death and division based on different rules produce vastly different outcomes in terms of the extent of segregation among the stiffer and softer cells (Figure 5(C)). As expected, the segregation parameter for the tissue without death and division is close to 1, signifying almost no segregation. The area, perimeter and random rule produce approximately the same segregation and the shape rule produces less[1,22].

The mixing of an initially segregated tissue shows strong differences in the outcomes for the different rules of cell-death and division, but the outcomes are also different from the initially mixed scenario (Figure 5(F)). Here, we look at the total length of the interface between the two types of cells to quantify the extent of interpenetration of the two cell types. A higher interfacial length corresponds to a higher degree of interpenetration. While tissue with perimeter-based rules for cell death and division had some of the highest degree of segregation, they also show the highest extent of interpenetration when starting from an initially segregated system. This increased interpenetration of the neighboring stiff cells might be driven by a combination of increased fluidity and increased proliferation of the softer cells in this system. The degree of interpenetration decreases with decreasing tissue fluidity and decreasing cell turnover rates.

In both the initially mixed and initially segregated scenarios, we also look at the ratio of the two types of cells and if the potential differences in proliferation and death rates between soft and stiff cells manifest differently during segregation or mixing. We expect the number of softer cells to be greater than the number of stiffer cells for death and division driven by the area, perimeter, and shape rules. This is because the energy corresponding to the cortical stiffness is given by $\frac{1}{2}k(P_i - P_0)^2$ (Equation 1), which allows for the softer cells to have a larger perimeter, consequently allowing for increased cell area and cell shape parameters as well, though weakly. These increases in cell parameters increase the likelihood of division across all the cell parameter-based rules, albeit with different strengths. However, it is interesting to note that the ratio of the cell types varies widely and is not a strong predictor of the segregation parameter or the interface length and that the death and division rule has a significant impact on the overall tissue level dynamics.

Lastly, the instantaneous velocity profiles of cells in these tissue systems highlight some interesting trends (Figure 5(B,E)). The cell death and division rules that increased tissue fluidization and cell turnover also increase overall cell speeds. More importantly, as seen in other studies, separated populations of softer cells showed high motility compared to separated populations of stiffer cells. This trend was observed irrespective of the various death and division rules, or an absence of death and division (later time points in Figure 5D, all time points in Figure 5E). However, when mixed, the motility of the softer cells influenced the motility of the stiffer cells and both populations moved at speeds comparable to those of separated softer cell populations. This matches observations seen in a previous experimental system[55]. An initially mixed system segregated into soft, fast-moving cells and stiff, slow-moving cells. However, when starting from an initially segregated system, even with interpenetration of the stiff cell regions by softer cells, there was no change in the average motility of either cell populations. We postulate that this is due to the dependence of the speed of a cell on the stiffness of the cells adjacent to it[3,56,57]. As de-mixing occurs in an initially mixed system, the stiffer cells are more likely to be surrounded by other stiffer cells, causing them to slow down[58]. This does not happen in the initially segregated system because cells are already surrounded by like neighbors and even with interpenetration, a large majority of the cells are still neighboring mechanically similar cells.

## Discussion

From our analysis we see that each of the rules produce differences in both microscopic properties such as speed of the cell and macroscopic properties such as invasiveness, but the most striking contrast is that all the tissues with any type of death and division have significantly different behavior as compared to the tissue without death and division. Thus, incorporating the effects of cell death and division is essential to truly understanding tissue organization and dynamics.

Cell death and division in any form increases average cell mobility. This is because cell death and division cause local decrease and increase in cell density respectively, leading to faster rearrangements of the cells to minimize the energy. The size-dependent rules of area and perimeter produce much more fluid tissues than their random and shape-dependent counterparts. We postulate that this is because of the density self-correcting nature of the size-dependent rules. If the local density of cells in any location of the tissue changes due to external stimulus or random perturbations, the likelihood of the cells to die or divide change rapidly under the size-dependent rule so as to bring the normalized cell density back to 1. The shape of a polyhedral cell is very weakly correlated to its size, so this effect is not very pronounced in the shape rule leading to lower fluidity. We have not considered changes in the active force generation ability of the cells in the tissue, which could be modeled by changing the effective $k_BT$ term in the Monte Carlo simulation, but we expect all tissues to become more fluid[59] as this internal energy is increased.

This fluidity does not however translate into invasiveness into an empty channel. We hypothesize that this difference is due to the lack of adhesion between the actual cells and the phantom cells in the channel. Due to this, the cells try to minimize their contact with the phantom cells. This results in a decrease in the size of cells entering the channel, potentially increasing their death rate. The shape rule being decoupled from the size of the cell is unaffected. Also, compression of the edge cells due to an unfavorable interface with the channel space can cause a flattening of these cells, driving proliferation based on the shape rule. Overall, the shape rule for death and division results in increased migration of the cells into the channels despite resulting in a less fluid, more jammed bulk system. The random rule's death and division dynamics are unaffected by local changes to the size of the cell and hence have the highest propensity to migrate into the channel.

In the de-mixing simulation setup, we see that any extent of death and division drives cells with different stiffnesses to de-mix from each other. We believe that this de-mixing is an outcome of three separate events – an initial unjamming transition due to cell death and division, and an increase in domain size due to division of cells, and an increase in cell division rate when surrounded by like-neighbors. Based on the extent to which each of these events contributes to the tissue reorganization, differences in segregation parameter, cell turnover rates and ratio of soft to stiff cells at the end of the simulation (Figure 5A-C, Supplementary Figure 3). It should be noted that a difference in the death and division rates between the two types of cells is not needed for de-mixing, as the random rule-based tissue also segregates. This suggests that the local slowing down (jamming) of stiff cells as they cluster post the initial unjamming transition and cell division events prevents any remixing of these populations.

The mixing simulation setup tells a similar story. The faster moving softer cells are unable to penetrate the slower moving stiffer cells in scenarios with no death and division or death and division driven by either random or shape-based death and division rules. In the scenarios with size-dependent death and division rules, the increased mobility and fluidity of the stiffer cell region allows for interpenetration by faster moving and faster proliferating softer cell populations. However, collective domains are still maintained and even with interpenetration, there is no direct mixing of the cell populations (Figures 5D-F, Supplementary Figure 4).

While the results described above qualitatively match prior predictions and observations in-terms of cell migration speeds for soft vs stiff cells, jammed vs unjammed nature of tissues with and without death and division, and mixing/de-mixing scenarios, direct validation at this point is difficult. However, the fact that a large variety of outcomes are observed in *in vivo* and *in vitro,* and there is significant complexity in fully describing tissue dynamics in healthy and diseased tissues, the specific results described here may point to key missing pieces of the overall puzzle.

This work considers death and division rules either primarily based on size or shape. The perimeter rule combines both the size and geometry and results in the most significant differences in outcomes compared to tissue dynamics without any cell death or division. Other rules combing different triggers for cellular apoptosis and division can also be conceived that could result in other emergent phenomena. Also, we have not considered clock-based rules of death and division where the age of a cell factors into its likelihood to die to divide. Our modeling framework can easily incorporate these additions and updates for future work.

## Conclusions

We have presented a first of its kind study describing the influence of various cell death and division triggering mechanisms on cell migration dynamics within tissues and overall tissue mechanics. We show that while cell death and division events indeed increase cell mobility within the tissues, this does not always translate to an overall decrease in tissue fluidity. Additionally, we observe complex interplays between tissue fluidity, tissue expansion propensity, tissue invasiveness, and self-organization within the tissue. These dynamics are strongly influenced by specific cell fate triggers. While the triggers we considered are limited to cell area, cell perimeter, cell shape or random, these triggers have strong biological relevance. Tissues that do not have any death and division events occurring behave significantly differently from tissues with even low cell turnover rates. In all, these results emphasize the need to accurately quantify the rules governing cell death and division within dense tissues to fully understand their dynamics. They also show the importance of incorporating different cell death and division mechanisms within in silico tissue models for improving their descriptive and predictive power.

## Conflicts of interest

The authors report no conflict of interest.

## Acknowledgements

We acknowledge Dr. George Youssef, San Diego State University for engaging with us in useful discussions regarding the analysis of tissue mechanics described here. P. K. Acknowledges financial support for this work under DoD ARO grant W911NF-17-1-0413.

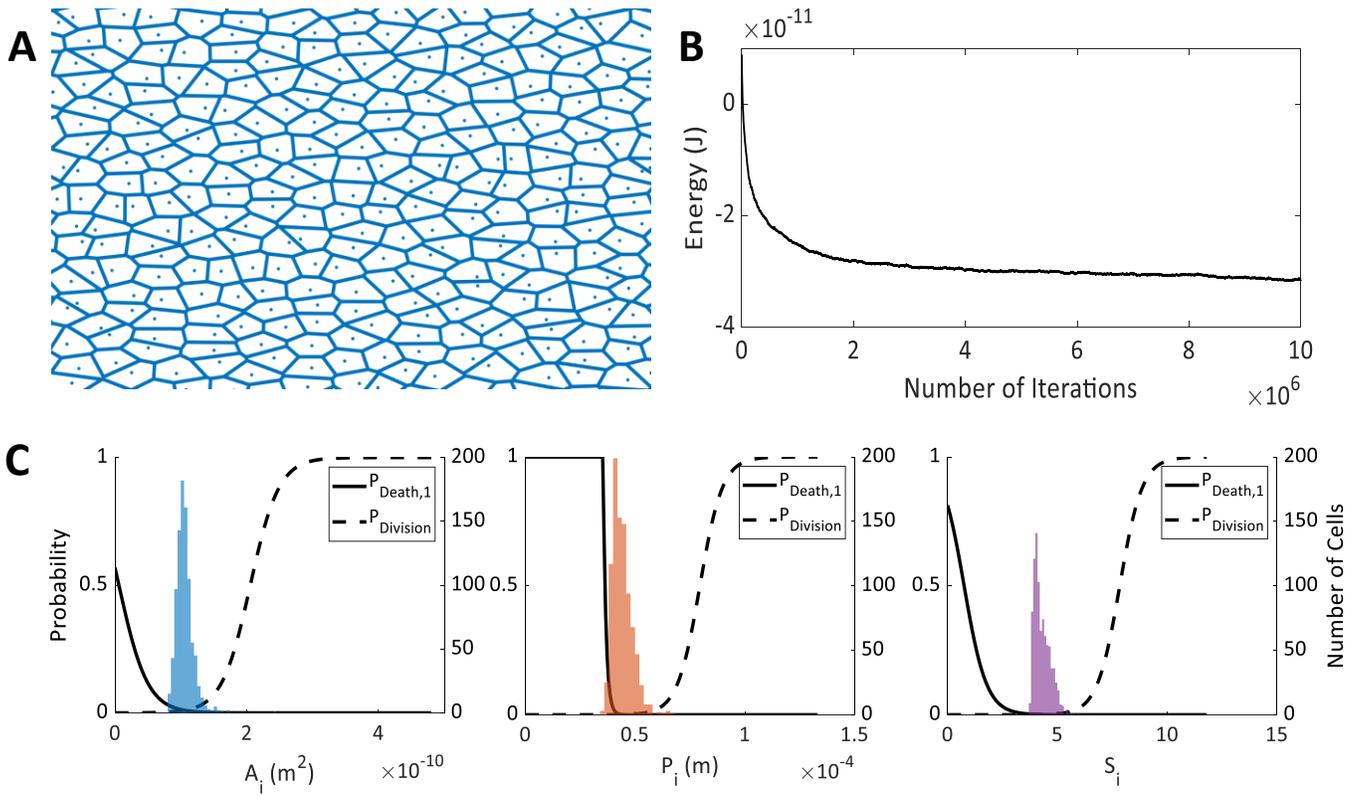

Figure 1: (A) Voronoi polygons representing cells where the dots are the nuclei (B) Energy minimization of the tissue during initial optimization (C) Distribution of area, perimeter, and shape of the cells respectively after initial optimization overlayed with cell death and division probabilities

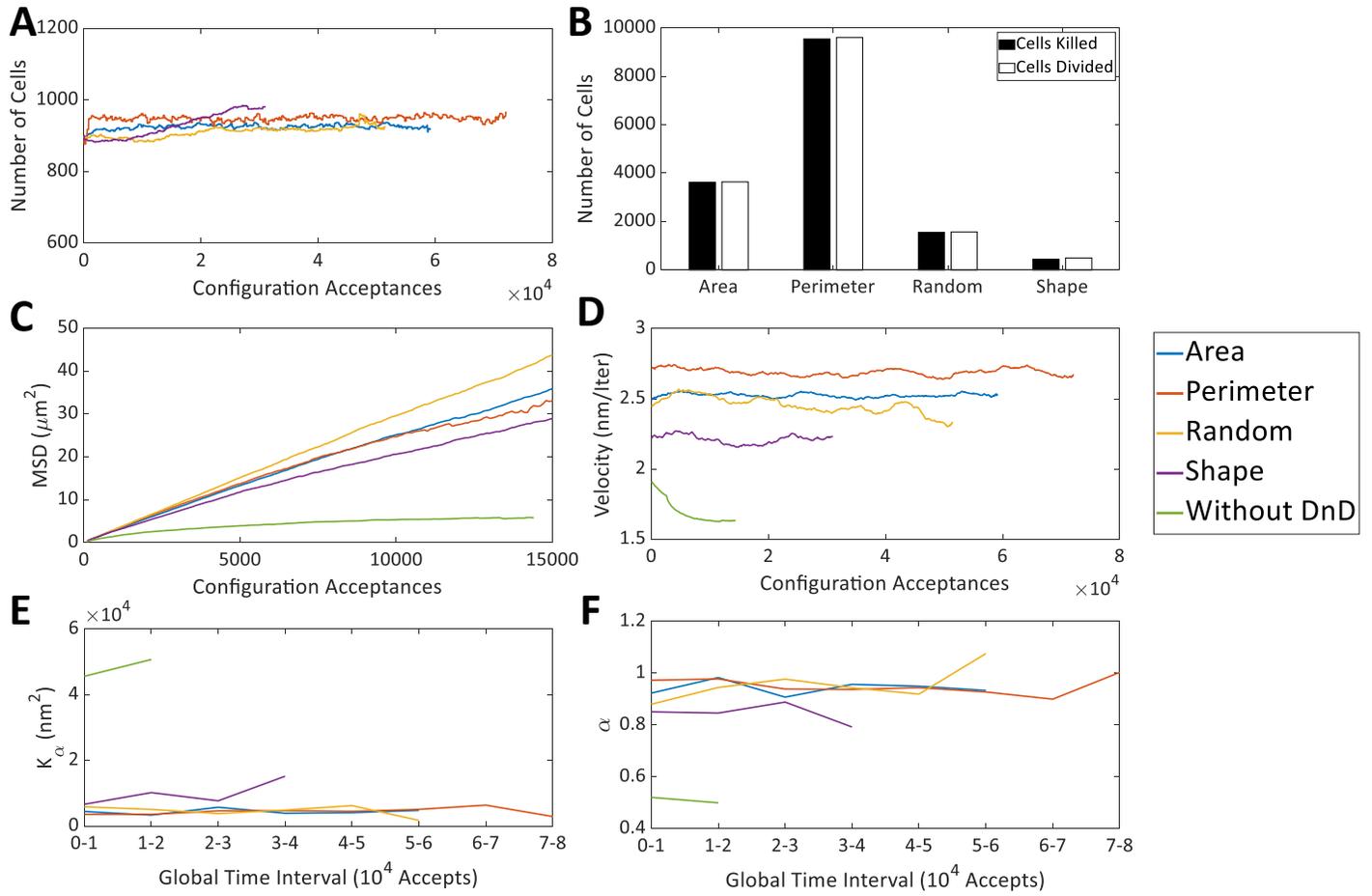

Figure 2: (A) Variation in number of cells against time (B) Total number of cells killed and divided for each of the rules, representing tissue turnover rate (C) Mean square of cell displacement (D) Instantaneous velocity (E-F) Variation of diffusion parameters $K_\alpha$ and $\alpha$ by global time interval

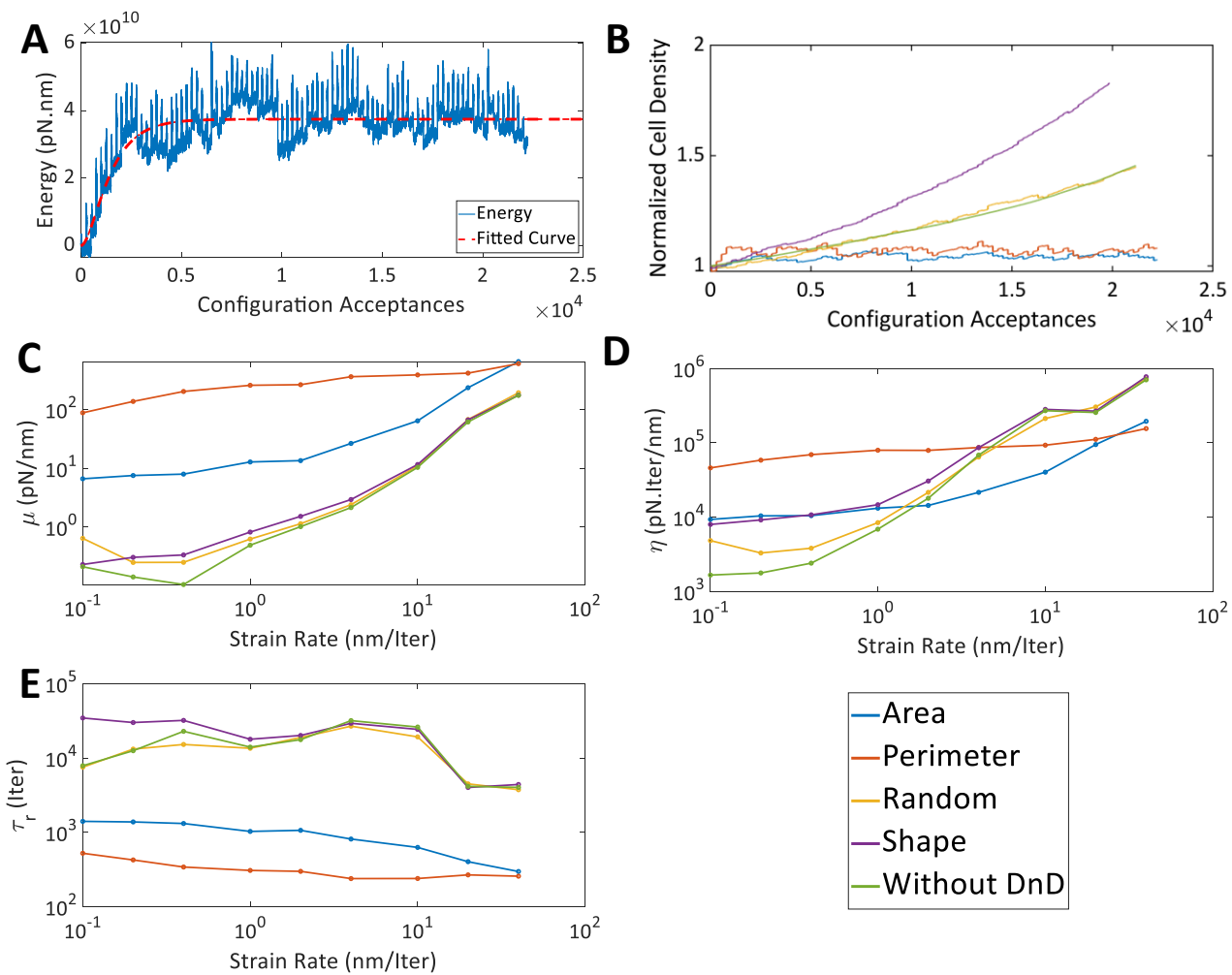

Figure 3: (A) Fitting of the energy profile with the Maxwell model for the area-based death and division rule (B) Variation of cell density during tissue compression (C-E) Viscoelastic response of tissues under different strain rates plotted on a log-log scale

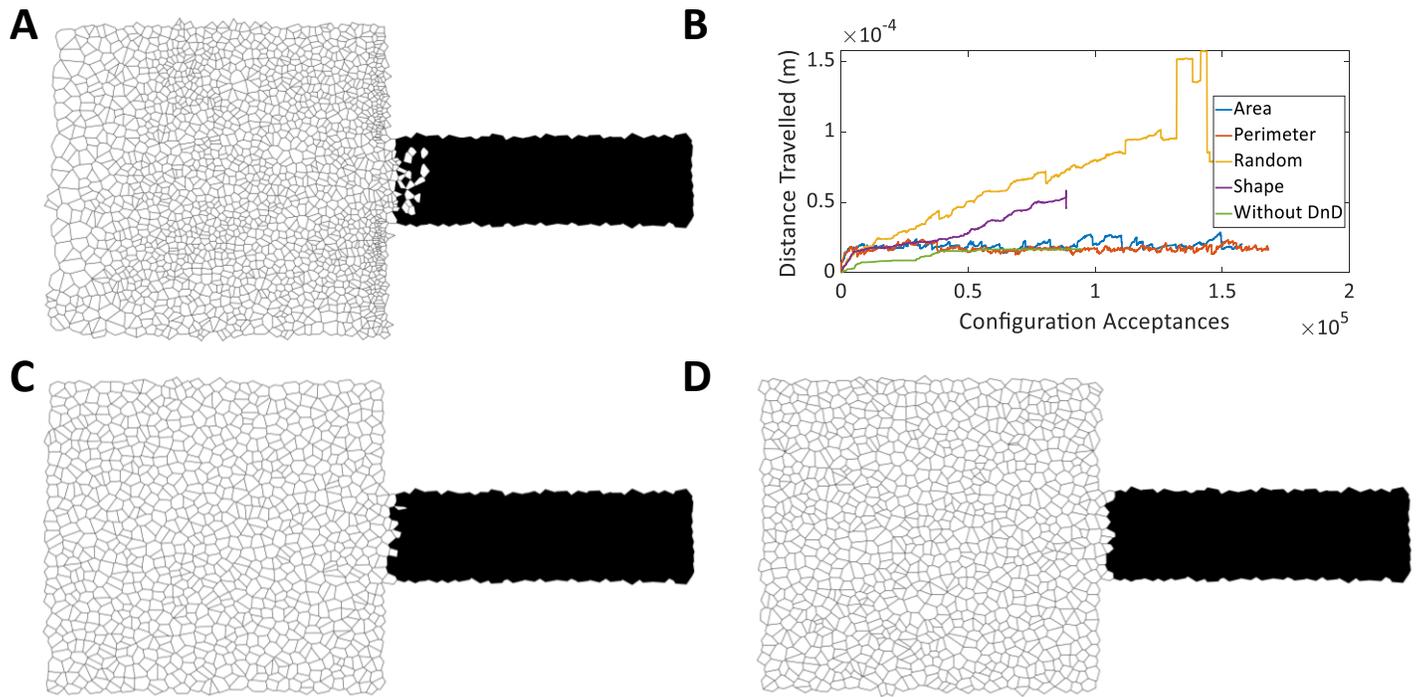

Figure 4: (A) Cells expanding into channel for the shape rule with the phantom cells marked in black (B) Distance travelled by leading cell into the channel for each of the death and division rules (C-D) Cells expanding into channel for the area and perimeter rules respectively

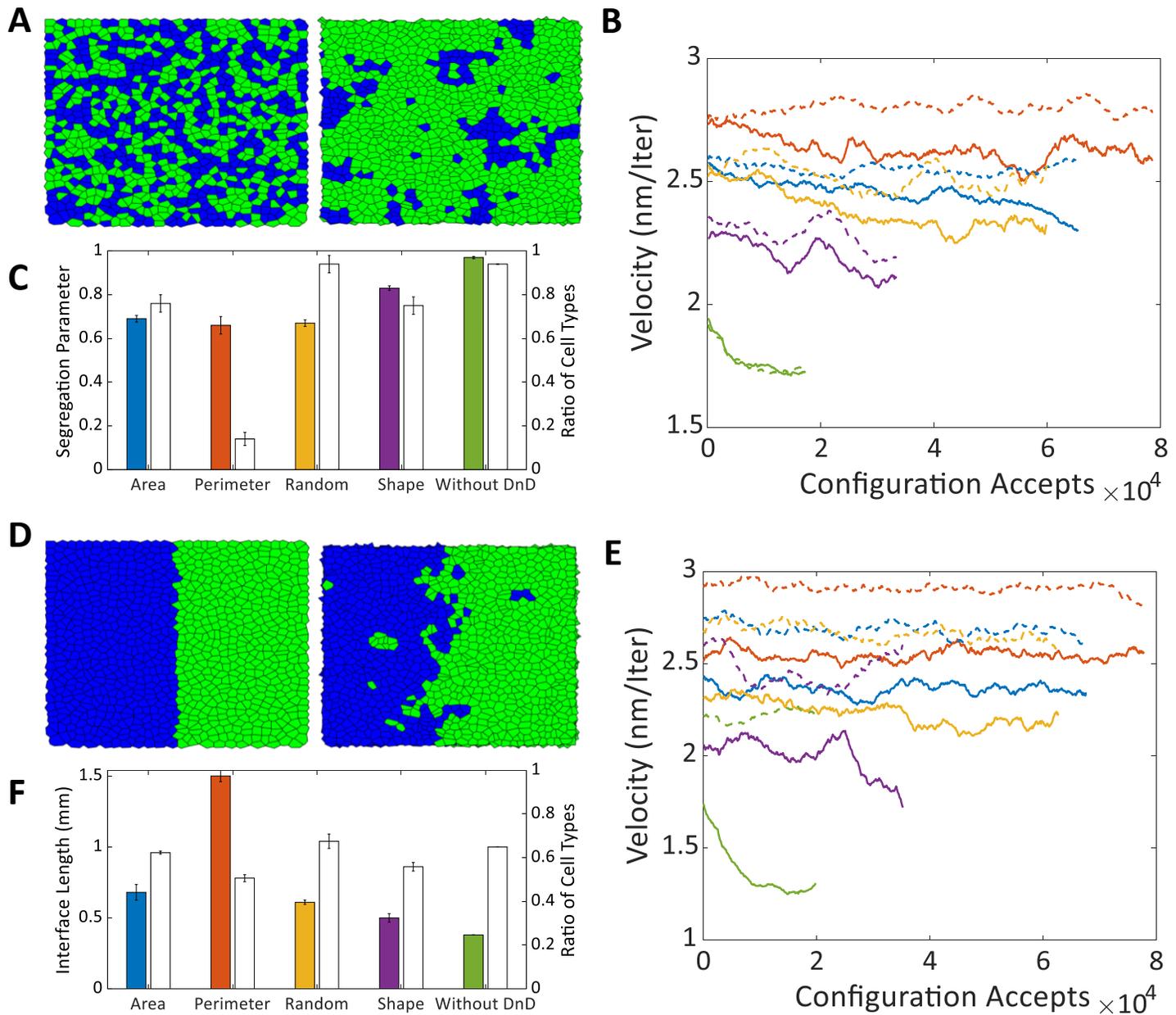

Figure 5: (A) De-mixing simulation setup with stiff (blue) and soft (green) cells; initial configuration (left) and final configuration for the perimeter rule (right) (B) Velocity profiles over time for stiff (solid line) and soft (dashed line) cells in the de-mixing setup (C) Segregation parameter and ratio of stiff to soft cells in the de-mixing setup (D) Mixing simulation setup with stiff (blue) and soft (green) cells; initial configuration (left) and final configuration for the perimeter rule (right) (E) Velocity profiles over time for stiff (solid line) and soft (dashed line) cells in the mixing setup (F) Length of interface between soft and stiff cells in the mixing setup

# Supplementary Materials

Supplementary Table 1: Obtained parameters from fitting the energy profile in the tissue compression simulation setup under each of the death and division rules for $\dot{\varepsilon} = 2 \, nm/Iter$ and cells with baseline stiffness (200 Pa). The cells highlighted in green are for the models which satisfy the fitting criteria.

| Model | Parameter | Death and Division Rule | | | | | |
|---|---|---|---|---|---|---|---|
| | | Area | 95% Confidence Interval | | Perimeter | 95% Confidence Interval | |
| 1 | $\mu$ | 8.218 | 8.097 | 8.339 | 324.7 | 314.8 | 334.6 |
| | $\eta$ | 1.165e+04 | 1.156e+04 | 1.173e+04 | 8.963e+04 | 8.827e+04 | 9.098e+04 |
| | $R^2$ | 0.8383 | | | 0.5751 | | |
| 2 | $\eta_1$ | 8.33e+04 | -2.414e+09 | 2.414e+09 | 1.57e+05 | -5.644e+09 | 5.644e+09 |
| | $\eta_2$ | 5.124e+05 | -3.211e+10 | 3.211e+10 | 1.18e+05 | -1.413e+10 | 1.413e+10 |
| | $\mu$ | 420.3 | -2.436e+07 | 2.436e+07 | 996.2 | -7.163e+07 | 7.163e+07 |
| | $R^2$ | 0.8383 | | | 0.5750 | | |
| 3 | $\mu_1$ | 8.218 | 8.097 | 8.339 | 324.7 | 314.8 | 334.6 |
| | $\mu_2$ | 8.268e-13 | fixed at bound | fixed at bound | 3.009e-14 | fixed at bound | fixed at bound |
| | $\eta$ | 1.165e+04 | 1.156e+04 | 1.173e+04 | 8.962e+04 | 8.827e+04 | 9.098e+04 |
| | $R^2$ | 0.8383 | | | 0.5751 | | |
| 4 | $\mu$ | 209.7 | -1.176e+07 | 1.176e+07 | 431.1 | -3.275e+07 | 3.275e+07 |
| | $\eta$ | 2.972e+05 | -1.666e+10 | 1.666e+10 | 1.19e+05 | -9.038e+09 | 9.038e+09 |
| | $F$ | -0.0004767 | -30.03 | 30.03 | -3.142e-05 | -10.23 | 10.23 |
| | $R^2$ | 0.8383 | | | 0.5750 | | |
| 5 | $\eta$ | 5.446e+05 | -3.585e+10 | 3.585e+10 | 2.253e+05 | -1.976e+10 | 1.976e+10 |
| | $\mu$ | 384.2 | -2.529e+07 | 2.529e+07 | 816.2 | -7.16e+07 | 7.16e+07 |
| | $F$ | 0.001249 | -76.94 | 76.94 | 0.0007348 | -51.99 | 51.99 |
| | $R^2$ | 0.8383 | | | 0.5750 | | |

| Model | Parameter | Random | 95% Confidence Interval | | Shape | 95% Confidence Interval | |
|---|---|---|---|---|---|---|---|
| 1 | $\mu$ | 0.7766 | 0.7739 | 0.7794 | 1.548 | 1.544 | 1.551 |
| | $\eta$ | 1.951e+04 | 1.946e+04 | 1.956e+04 | 2.897e+04 | 2.895e+04 | 2.9e+04 |
| | $R^2$ | 0.9936 | | | 0.9979 | | |
| 2 | $\eta_1$ | 1.16e+05 | -8.44e+08 | 8.442e+08 | 1.447e+05 | -6.71e+08 | 6.712e+08 |
| | $\eta_2$ | 5.733e+05 | -9.189e+09 | 9.19e+09 | 5.78e+05 | -6.032e+09 | 6.033e+09 |
| | $\mu$ | 27.44 | -3.993e+05 | 3.994e+05 | 38.6 | -3.58e+05 | 3.581e+05 |
| | $R^2$ | 0.9936 | | | 0.9979 | | |
| 3 | $\mu_1$ | 2.249 | 2.219 | 2.278 | 1.548 | 1.544 | 1.551 |
| | $\mu_2$ | 0.3659 | 0.3653 | 0.3665 | 4.803e-09 | fixed at bound | fixed at bound |
| | $\eta$ | 9412 | 9392 | 9433 | 2.897e+04 | 2.895e+04 | 2.9e+04 |
| | $R^2$ | 0.9974 | | | 0.9979 | | |
| 4 | $\mu$ | 23.59 | -4.143e+05 | 4.143e+05 | 33.03 | -3.733e+05 | 3.734e+05 |
| | $\eta$ | 5.927e+05 | -1.041e+10 | 1.041e+10 | 6.184e+05 | -6.989e+09 | 6.99e+09 |
| | $F$ | -0.0009703 | -18.93 | 18.93 | -0.0009691 | -12.47 | 12.47 |
| | $R^2$ | 0.9936 | | | 0.9979 | | |
| 5 | $\eta$ | 6.396e+05 | -1.129e+10 | 1.129e+10 | 5.494e+05 | -6.262e+09 | 6.263e+09 |
| | $\mu$ | 25.46 | -4.492e+05 | 4.493e+05 | 29.35 | -3.345e+05 | 3.346e+05 |
| | $F$ | 0.001503 | -24.54 | 24.55 | 0.001351 | -13.96 | 13.96 |

| | $R^2$ | 0.9936 | | | 0.9979 | | |
|---|---|---|---|---|---|---|---|

Supplementary Table 2: Obtained parameters from fitting the energy profile in the tissue compression simulation setup under each of the death and division rules for $\dot{\varepsilon} = 2 \; nm/Iter$ and cells with half stiffness (100 Pa). The cells highlighted in green are for the models which satisfy the fitting criteria.

| | | Death and Division Rule | | | | | |
|---|---|---|---|---|---|---|---|
| Model | Parameter | Area | 95% Confidence Interval | | Perimeter | 95% Confidence Interval | |
| 1 | $\mu$ | 13.59 | 13.32 | 13.87 | 257.4 | 252.1 | 262.7 |
| | $\eta$ | 1.387e+04 | 1.373e+04 | 1.401e+04 | 8.351e+04 | 8.265e+04 | 8.436e+04 |
| | $R^2$ | 0.7169 | | | 0.7275 | | |
| 2 | $\eta_1$ | 1.354e+05 | -4.771e+09 | 4.771e+09 | 2.466e+05 | -7.97e+09 | 7.97e+09 |
| | $\eta_2$ | 1.186e+06 | -8.835e+10 | 8.835e+10 | 4.818e+05 | -3.911e+10 | 3.911e+10 |
| | $\mu$ | 1295 | -9.126e+07 | 9.126e+07 | 2245 | -1.451e+08 | 1.451e+08 |
| | $R^2$ | 0.7169 | | | 0.7275 | | |
| 3 | $\mu_1$ | 13.59 | 13.32 | 13.87 | 257.9 | 252.5 | 263.2 |
| | $\mu_2$ | 2.904e-12 | fixed at bound | fixed at bound | 9.788e-05 | -7.959e-05 | 0.0002754 |
| | $\eta$ | 1.387e+04 | 1.373e+04 | 1.401e+04 | 8.356e+04 | 8.27e+04 | 8.442e+04 |
| | $R^2$ | 0.7169 | | | 0.7275 | | |
| 4 | $\mu$ | 591.3 | -4.238e+07 | 4.238e+07 | 915.3 | -5.175e+07 | 5.176e+07 |
| | $\eta$ | 6.034e+05 | -4.324e+10 | 4.324e+10 | 2.97e+05 | -1.679e+10 | 1.679e+10 |
| | $F$ | -0.001024 | -79.93 | 79.93 | -0.000279 | -24.68 | 24.68 |
| | $R^2$ | 0.7169 | | | 0.7275 | | |
| 5 | $\eta$ | 6.251e+05 | -4.49e+10 | 4.49e+10 | 2.298e+05 | -1.55e+10 | 1.55e+10 |
| | $\mu$ | 612.6 | -4.4e+07 | 4.4e+07 | 708.4 | -4.779e+07 | 4.779e+07 |
| | $F$ | 0.001436 | -96.48 | 96.48 | 0.0007368 | -40.35 | 40.35 |
| | $R^2$ | 0.7169 | | | 0.7275 | | |

| Model | Parameter | Random | 95% Confidence Interval | | Shape | 95% Confidence Interval | |
|---|---|---|---|---|---|---|---|
| 1 | $\mu$ | 1.017 | 1.015 | 1.02 | 1.338 | 1.335 | 1.342 |
| | $\eta$ | 2.112e+04 | 2.109e+04 | 2.115e+04 | 3.025e+04 | 3.02e+04 | 3.03e+04 |
| | $R^2$ | 0.9971 | | | 0.9969 | | |
| 2 | $\eta_1$ | 1.698e+05 | -9.448e+08 | 9.452e+08 | 2.063e+05 | -1.189e+09 | 1.189e+09 |
| | $\eta_2$ | 1.196e+06 | -1.425e+10 | 1.426e+10 | 1.201e+06 | -1.503e+10 | 1.503e+10 |
| | $\mu$ | 65.79 | -7.32e+05 | 7.322e+05 | 62.25 | -7.173e+05 | 7.174e+05 |
| | $R^2$ | 0.9971 | | | 0.9969 | | |
| 3 | $\mu_1$ | 1.017 | 1.015 | 1.02 | 1.338 | 1.335 | 1.342 |
| | $\mu_2$ | 2.412e-10 | fixed at bound | fixed at bound | 5.618e-09 | fixed at bound | fixed at bound |
| | $\eta$ | 2.112e+04 | 2.109e+04 | 2.115e+04 | 3.025e+04 | 3.02e+04 | 3.03e+04 |
| | $R^2$ | 0.9971 | | | 0.9969 | | |
| 4 | $\mu$ | 9.424 | -2.267e+05 | 2.267e+05 | 24.52 | -3.298e+05 | 3.299e+05 |
| | $\eta$ | 5.369e+05 | -1.291e+10 | 1.291e+10 | 5.544e+05 | -7.456e+09 | 7.458e+09 |
| | $F$ | -0.0007992 | -22.53 | 22.53 | -0.0008498 | -13.17 | 13.17 |
| | $R^2$ | 0.9727 | | | 0.9969 | | |
| 5 | $\eta$ | 5.432e+05 | -7.022e+09 | 7.023e+09 | 6.026e+05 | -8.074e+09 | 8.075e+09 |
| | $\mu$ | 26.16 | -3.382e+05 | 3.383e+05 | 26.66 | -3.572e+05 | 3.572e+05 |

| | F | 0.001301 | -15.43 | 15.43 | 0.001475 | -17.96 | 17.96 |
|---|---|---|---|---|---|---|---|
| | $R^2$ | 0.9971 | | | 0.9969 | | |

Supplementary Table 3: Derivation of energy expression in the models

In all the models, the strain is being applied at point A towards the right while point C is stationary (Fixed to ground). We assume that points A and B move a distance of $x_1$ and $x_2$ respectively towards the right with respect to point C. The energy of the system is assumed to be the energy stored in the spring(s), since it is the only restorative force. Since the point A is moving at a constant strain rate, for all the models we have –

$$\frac{dx_1}{dt} = \dot{\varepsilon} \Rightarrow x_1 = \dot{\varepsilon}\, t$$

Lastly, we balance the forces at point B to get the relationship between $x_1$ and $x_2$ and derive the energy stored in the spring.

| | Model Diagram | Derivation of Energy Expression |
|---|---|---|
| 1 | 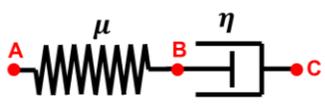 | Force balance at point B – $$\mu(x_1 - x_2) = \eta \frac{dx_2}{dt}$$ $$\Rightarrow \eta \frac{dx_2}{dt} + \mu x_2 = \mu \dot{\varepsilon}\, t$$ Solving ODE using initial condition that $x_1 = x_2 = 0$ at $t = 0$ $$x_2 = \frac{\eta \dot{\varepsilon}}{\mu}\left(e^{\frac{-\mu t}{\eta}} - 1\right) + \dot{\varepsilon}\, t$$ Plugging into energy expression – $$E = \frac{\mu}{2}(x_2 - x_1)^2 = \frac{\mu}{2}(x_2 - \dot{\varepsilon}\, t)^2$$ $$\Rightarrow E = \frac{\mu}{2}\left(\frac{\eta\dot{\varepsilon}}{\mu}(e^{\frac{-\mu t}{\eta}} - 1)\right)^2$$ |
| 2 | 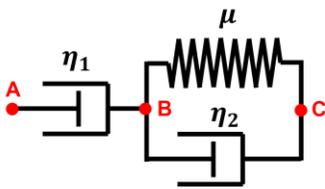 | Force balance at point B – $$\eta_1\left(\frac{dx_1}{dt} - \frac{dx_2}{dt}\right) = \eta_2 \frac{dx_2}{dt} + \mu x_2$$ $$\Rightarrow (\eta_1 + \eta_2)\frac{dx_2}{dt} + \mu x_2 = \eta_1 \dot{\varepsilon}$$ Solving ODE using initial condition that $x_1 = x_2 = 0$ at $t = 0$ $$x_2 = \frac{\eta_1 \dot{\varepsilon}}{\mu}\left(1 - e^{\frac{-\mu t}{\eta_1 + \eta_2}}\right)$$ Plugging into energy expression – $$E = \frac{\mu}{2}(x_2)^2$$ $$\Rightarrow E = \frac{\mu}{2}\left(\frac{\eta_1\dot{\varepsilon}}{\mu}(1 - e^{\frac{-\mu t}{\eta_1+\eta_2}})\right)^2$$ |
| 3 | 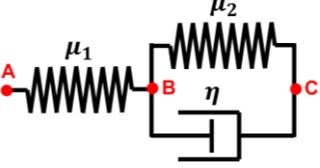 | Force balance at point B – $$\mu_1(x_1 - x_2) = \eta \frac{dx_2}{dt} + \mu_2 x_2$$ $$\Rightarrow \eta \frac{dx_2}{dt} + (\mu_1 + \mu_2)x_2 = \mu_1 \dot{\varepsilon} t$$ Solving ODE using initial condition that $x_1 = x_2 = 0$ at $t = 0$ $$x_2 = \frac{\eta \mu_1 \dot{\varepsilon}}{(\mu_1 + \mu_2)^2}\left(e^{\frac{-(\mu_1+\mu_2)t}{\eta}} - 1\right) + \frac{\mu_1 \dot{\varepsilon} t}{\mu_1 + \mu_2}$$ Plugging into energy expression – $$E = \frac{\mu_1}{2}(x_2 - x_1)^2 + \frac{\mu_2}{2}(x_2)^2$$ $$\Rightarrow E = \frac{\mu_1}{2}(\dot{\varepsilon}t - \frac{\eta\mu_1\dot{\varepsilon}}{(\mu_1+\mu_2)^2}\left(e^{\frac{-(\mu_1+\mu_2)t}{\eta}} - 1\right) - \frac{\mu_1\dot{\varepsilon}t}{\mu_1+\mu_2})^2$$ $$+ \frac{\mu_2}{2}(\frac{\eta\mu_1\dot{\varepsilon}}{(\mu_1+\mu_2)^2}\left(e^{\frac{-(\mu_1+\mu_2)t}{\eta}} - 1\right) + \frac{\mu_1\dot{\varepsilon}t}{\mu_1+\mu_2})^2$$ |
| 4 | 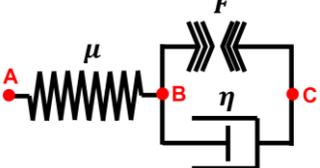 | Force balance at point B – $$\mu(x_1 - x_2) = \eta \frac{dx_2}{dt} + F$$ $$\Rightarrow \eta \frac{dx_2}{dt} + \mu x_2 = \mu \dot{\varepsilon} t - F$$ Solving ODE using initial condition that $x_1 = x_2 = 0$ at $t = 0$ $$x_2 = \left(\frac{\eta\dot{\varepsilon}}{\mu} + \frac{F}{\mu}\right)\left(e^{\frac{-\mu t}{\eta}} - 1\right) + \dot{\varepsilon}t$$ Plugging into energy expression – $$E = \frac{\mu}{2}(x_2 - x_1)^2$$ $$\Rightarrow E = \frac{\mu}{2}((\frac{\eta\dot{\varepsilon}}{\mu} + \frac{F}{\mu})(e^{\frac{-\mu t}{\eta}} - 1))^2$$ |
| 5 | 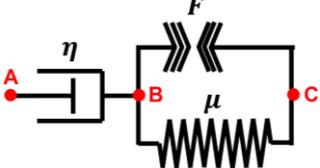 | Force balance at point B – $$\eta\left(\frac{dx_1}{dt} - \frac{dx_2}{dt}\right) = \mu x_2 + F$$ $$\Rightarrow \eta \frac{dx_2}{dt} + \mu x_2 = \eta \dot{\varepsilon} - F$$ Solving ODE using initial condition that $x_1 = x_2 = 0$ at $t = 0$ $$x_2 = \left(\frac{\eta\dot{\varepsilon}}{\mu} - \frac{F}{\mu}\right)\left(1 - e^{\frac{-\mu t}{\eta}}\right)$$ Plugging into energy expression – $$E = \frac{\mu}{2}(x_2)^2$$ $$\Rightarrow E = \frac{\mu}{2}((\frac{\eta\dot{\varepsilon}}{\mu} - \frac{F}{\mu})(1 - e^{\frac{-\mu t}{\eta}}))^2$$ |

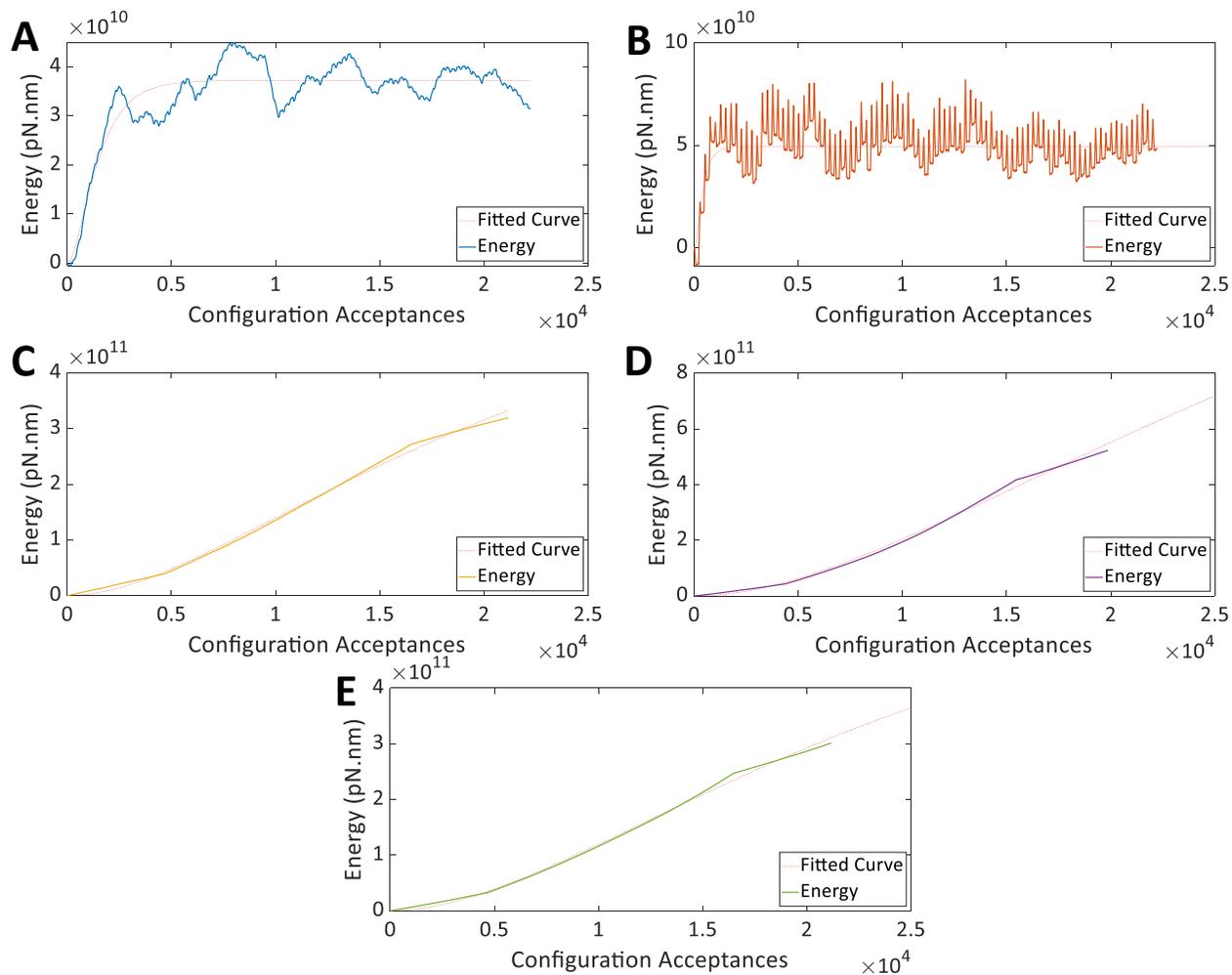

Supplementary Figure 1: (A-E) Fitting the energy profiles of the tissue compression scenario under the area, perimeter, random, shape and without death and division rules respectively

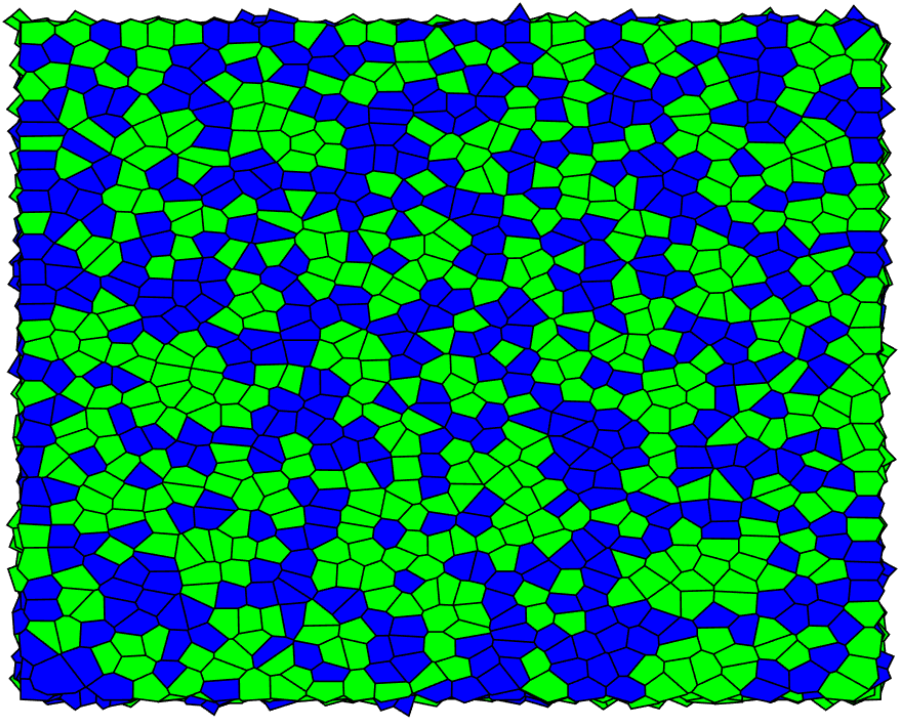

Supplementary Figure 2: Final configuration of an initially mixed system of cells without death and division

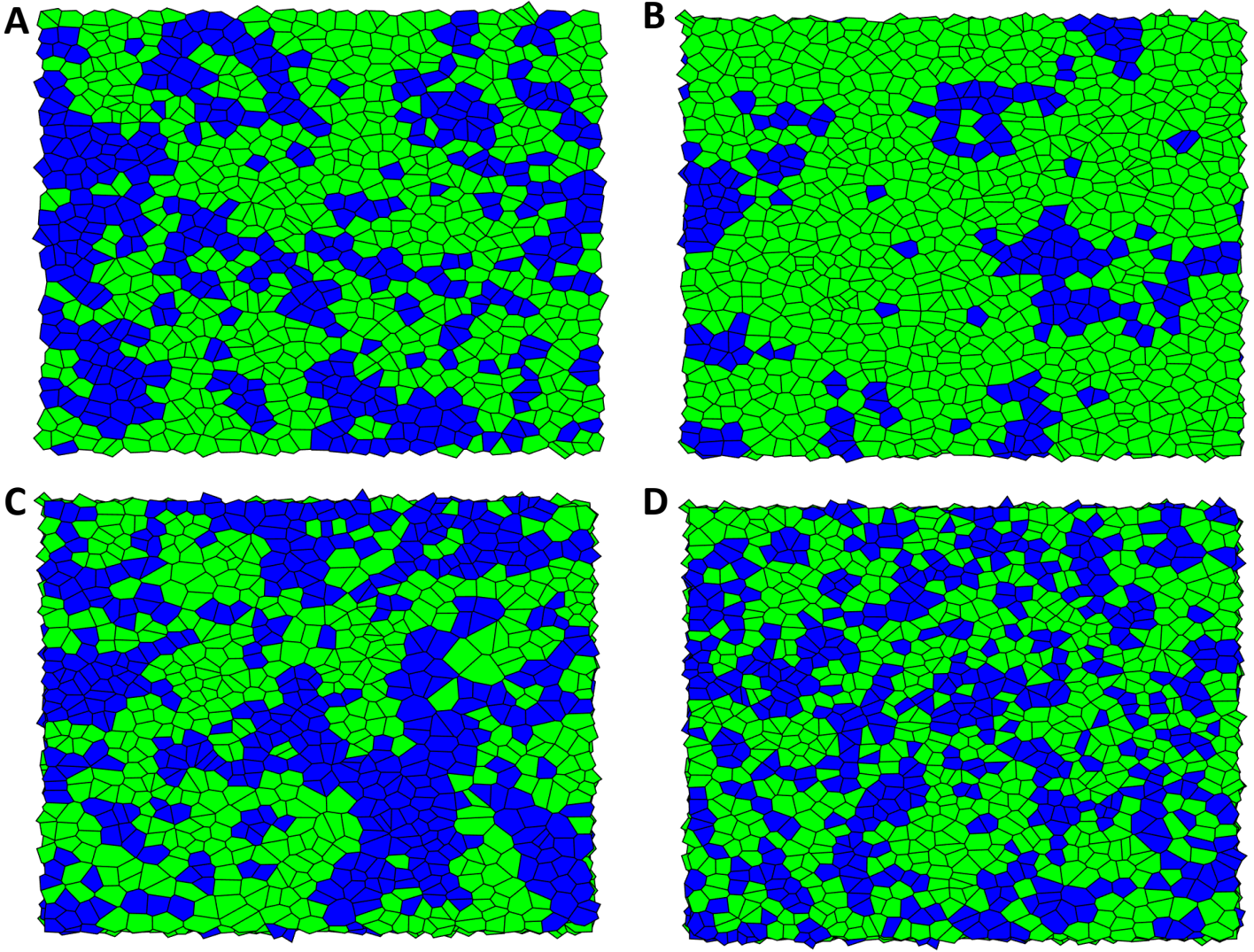

Supplementary Figure 3: (A-D) Final configuration of an initially mixed system of cells under the area, perimeter, random and shape rule respectively

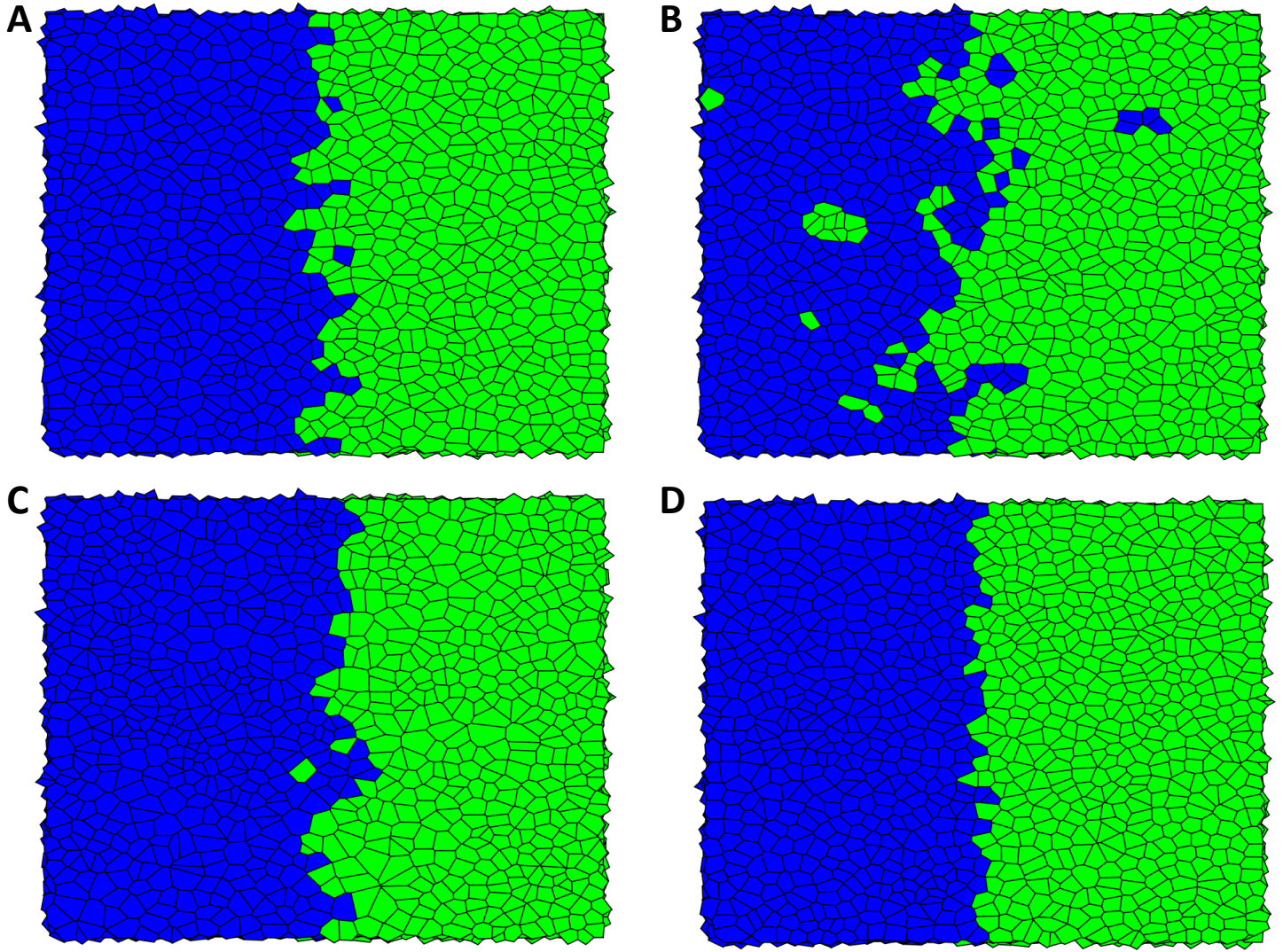

Supplementary Figure 4: (A-D) Final configuration of an initially segregated system of cells under the area, perimeter, random and shape rule respectively